\documentclass[twocolumn,showpacs,preprintnumbers,amsmath,amssymb,prb,superscriptaddress]{revtex4}
\usepackage{graphicx}
\usepackage{dcolumn}
\usepackage{bm}
\begin{document}
\title{Antiferromagnetism versus Kondo screening in the two-dimensional 
periodic Anderson model at half filling: Variational cluster approach}
\author{S. Horiuchi}
\affiliation{Department of Physics, Chiba University, Inage-ku, Chiba 263-8522, Japan}
\author{S. Kudo}
\affiliation{Department of Physics, Chiba University, Inage-ku, Chiba 263-8522, Japan}
\author{T. Shirakawa}
\affiliation{Department of Physics, Chiba University, Inage-ku, Chiba 263-8522, Japan}
\affiliation{Institut f\"ur Theoretische Physik, Leibnitz Universit\"at 
Hannover, D-30167 Hannover, Germany}
\author{Y. Ohta}
\affiliation{Department of Physics, Chiba University, Inage-ku, Chiba 263-8522, Japan}
\date{\today}

\begin{abstract}
The variational cluster approach (VCA) based on the self-energy 
functional theory is applied to the two-dimensional symmetric periodic 
Anderson model at half filling.  
We calculate a variety of physical quantities including the staggered 
moments and single-particle spectra at zero temperature to show that 
the symmetry breaking due to antiferromagnetic ordering occurs in the 
strong coupling region, whereas in the weak coupling region, the Kondo 
insulating state without symmetry breaking is realized.  The critical 
interaction strength is estimated.  
We thus demonstrate that the phase transition due to competition 
between antiferromagnetism and Kondo screening in the model can be 
described quantitatively by VCA.  
\end{abstract}

\maketitle

\section{\label{sec:Intro}Introduction}

The two-dimensional (2D) heavy fermion system has attracted 
much attention as a subsequent study of high-temperature 
superconductors and has recently been one of the central 
issues in the study of strongly correlated electron systems.  
For example, the heavy fermion material YbRh$_2$Si$_2$ shows 
a rapid change in the Hall coefficient as a function of magnetic 
field at zero temperature, which is accompanied by the 
antiferromagnetic (AF) to paramagnetic (PM) phase 
transition.\cite{paschen}  Also, heavy-fermion--like behavior 
is observed in the system of $^3$He bi-layers adsorbed on 
graphite.\cite{neumann}  
Generally speaking, competition between the magnetic ordering 
of localized spins through the Ruderman-Kittel-Kasuya-Yoshida 
(RKKY)\cite{kittel} interaction and the nonmagnetic states induced 
by the Kondo screening\cite{kondo} brings about the observed 
anomalous behaviors in heavy fermion materials.  
From the theoretical point of view, the periodic Anderson 
model (PAM)\cite{anderson} is one of the simplified models 
for heavy fermion systems, which is believed to describe the 
competition between magnetic ordering and Kondo singlet formation 
observed in heavy fermion materials.  

In 2D quantum systems, the symmetry-broken magnetically ordered 
state can be realized in the ground state at zero temperature and 
therefore one needs a method of calculation appropriate for 
infinite-size systems in the thermodynamic limit.  
In this paper, we therefore use the variational cluster approach 
(VCA)\cite{potthoffPRL,dahnken} based on the self-energy 
functional theory (SFT)\cite{potthoff} to consider the 2D periodic 
Anderson model at half filling.  
Although the self-energies of the small clusters are used in the 
VCA calculations and thus the long-range spin fluctuations beyond 
the cluster size are not taken into account, the quantum fluctuations 
within the cluster are treated exactly in this approach.  We may 
therefore expect that this approach should be applicable to describe 
the possible symmetry breaking of the model in the thermodynamic 
limit beyond the simple mean-field theory.  
We want to point out that the present calculation is the first one 
where the VCA is applied to the 2D PAM, as far as we know.  

We will show that, by means of VCA, the symmetry breaking due to the 
AF ordering of localized spins occurs in the strong coupling 
region, whereas in the weak coupling region, the Kondo insulator 
without symmetry breaking is realized.  The critical interaction 
strength will thereby be determined.  
We will also calculate the staggered magnetic moment as a function 
of the interaction strength and show that the phase transition is 
of the second order.  
We will furthermore calculate the single-particle spectra and 
densities of states (DOS) to discuss the effects of electron correlation 
on the quasiparticle band structure.  We will thus show how the system 
changes from the AF insulator, Kondo insulator, to the band insulator, 
with decreasing the interaction strength.  

This paper is organized as follows. 
In Sec.~\ref{sec:Method}, we present our model and method of 
calculation.  In Sec.~\ref{sec:Results}, we present our results 
of calculations for the stability of the AF ordering, staggered magnetic 
moment, single-particle spectra, and DOS by VCA.  
We summarize our work in Sec.~\ref{sec:Summary}.  

 \section{\label{sec:Method}Model and Method}

\subsection{\label{Model}Model}

We consider the PAM defined on the 2D square lattice.  
The Hamiltonian is given by 
\begin{eqnarray}
H&=&-t\sum_{\langle ij\rangle}
(c_{i\sigma}^{\dagger}c_{j\sigma}+{\rm H.c.})
-V\sum_{i\sigma}
(c_{i\sigma}^{\dagger}f_{i\sigma}+{\rm H.c.})\nonumber \\
&&+U\sum_{i}n_{i\uparrow}^fn_{i\downarrow}^f
+\varepsilon_f\sum_{i\sigma}n_{i\sigma}^f ,
\end{eqnarray}
where $c_{i\sigma}$ $(f_{i\sigma})$ is the annihilation 
operator of an electron at site $i$ with spin $\sigma$ 
in the conduction-electron $c$ ($f$-electron $f$) orbital, 
and $n_{i\sigma}^f=f_{i\sigma}^{\dagger}f_{i\sigma}$ 
is the electron number operator in the $f$ orbital.  
$t$ is the hopping parameter between the nearest-neighbor 
$c$ orbitals, $V$ is the on-site hybridization parameter between 
the $c$ and $f$ orbitals, $U$ is the on-site repulsion on 
the $f$ orbital, and $\varepsilon_f$ is the energy level of 
the $f$ orbital with respect to that of the $c$ orbital set 
to be the origin of energy.  
In the following calculations, we consider the symmetric case, 
i.e., the case with $\varepsilon_f=-U/2$.  We also focus on 
the electron densities at half filling, i.e., $2N_s$ electrons 
in the $N_s$ unit cells, where the unit cell contains one $c$ 
and one $f$ orbital.  
We hereafter set $t=V=1$ as the unit of energy and we change 
the value of the interaction strength $U$.  

\subsection{\label{VCA}Variational cluster approach}

Let us first briefly review the formulation of 
SFT\cite{potthoff} and present the method of calculation 
of the magnetic ordering by VCA\cite{potthoffPRL,dahnken} 
in order to make our paper self-contained.  

We consider the system of the Hamiltonian 
$H=H_0({\bm t})+H_1({\bm U})$, 
where ${\bm t}$ and ${\bm U}$ denote the one-particle 
and interaction parameters of $H$, respectively. 
In general, the grand potential is given from 
the stationary point of the self-energy functional 
\begin{equation}
\Omega[\Sigma]=F[\Sigma]+{\rm Tr}\ln[-(G_0^{-1}-\Sigma)^{-1}],
\end{equation}
where $F[\Sigma]$ and $G_0$ are the Legendre transform of 
the Luttinger-Ward potential $\Phi[G]$ and the bare Green function, 
respectively.  The rigorous variational principle 
$\delta\Omega[\Sigma]/\delta\Sigma=0$ gives the 
Dyson equation $G^{-1}=G_0^{-1}-\Sigma$, where 
$G$ is the physical Green function.  

In the above expression (2), $F[\Sigma]$ is a universal functional  
of the self-energy; i.e., $F[\Sigma]$ remains unchanged for an 
arbitrary reference system of the Hamiltonian 
$H'=H_0({\bm t'})+H_1({\bm U})$ 
that has the same interaction part as the original system has, 
but with modified one-particle parameters.  
We here introduce the restriction of the space of the 
exact self-energies of the original system to the set of 
exact self-energies of the reference system.  
Because of this restriction, the following procedure 
becomes approximate but it enables us to obtain the 
grand potential  of the original system from the 
stationary point of the $\Sigma({\bm t'})$ functional 
\begin{eqnarray}
\Omega[\Sigma({\bm t'})]=\Omega'
&+&{\rm Tr}\ln[-(G_0^{-1}-\Sigma({\bm t'}))^{-1}] \nonumber \\
&-&{\rm Tr}\ln[-({G'}_0^{-1}-\Sigma({\bm t'}))^{-1}],
\label{eq:omega}
\end{eqnarray}
where $\Omega'$, $\Sigma({\bm t'})$, and ${G'}_0$ are 
the grand potential, exact self-energy, and bare Green function 
of the reference system, respectively. 
The condition 
$\partial\Omega[\Sigma({\bm t'})]/\partial{\bm t'}=0$ 
gives an appropriate reference system that describes the 
original system approximately. 

In VCA, we first divide the original infinite lattice into 
the finite-size identical clusters.  
By switching off the hopping parameters between the identical 
clusters, we construct the reference system as an assembly 
of the exactly solvable finite-site clusters.  
One of the major advantages of VCA is its ability to describe 
the symmetry-breaking long-range order by introducing suitably 
chosen fictitious symmetry-breaking Weiss fields in the set 
of variational parameters ${\bm t'}$. 
In order to discuss the competition between the AF ordering 
and Kondo screening in the parameter space, we here introduce 
staggered magnetic field $h'$ on the $f$ orbitals in the cluster 
Hamiltonian as a variational parameter.  We thus obtain the 
Hamiltonian of the reference system, $H'$, which is given by 
\begin{eqnarray}
H'&=&\sum_{\bm R}H'_{\bm R}, \\
H'_{\bm R}&=&-t\sum_{\langle ij\rangle}
(c_{i\sigma}^{\dagger}c_{j\sigma}+{\rm H.c.})
-V\sum_{i\sigma}
(c_{i\sigma}^{\dagger}f_{i\sigma}+{\rm H.c.})\nonumber \\
&&+U\sum_{i}n_{i\uparrow}^fn_{i\downarrow}^f
+\varepsilon_f\sum_{i\sigma}n_{i\sigma}^f \nonumber \\
&&+h'\sum_i e^{i{\bm Q}\cdot{\bm r}_i}(n_{i\uparrow}^f-n_{i\downarrow}^f),
\end{eqnarray}
where ${\bm R}$ is the label of the clusters, $i$ and $j$ are 
the labels of the sites within the cluster ${\bm R}$, and 
${\bm Q}=(\pi,\pi)$.  

In the present study, we use a 6-site ($2\times 3$) cluster 
to search for the stationary point of $\Omega[\Sigma(h')]$ 
with a condition $\partial\Omega[\Sigma(h')]/\partial h'=0$ 
as discussed above.  
We should note that the shape of the cluster introduced as 
a reference system is not commensurate with the AF ordering.  
We therefore treat a 12-site ($2\times 6$) cluster as a 
supercell by combining the two 6-site clusters.  
We treat the intercluster hopping elements as well as the 
hopping elements between the supercells 
``perturbatively'';\cite{aichhorn} 
i.e., we use the self-energies of the 6-site clusters to 
calculate the Green function of the original infinite system 
as well as that of the reference systems (an assembly of the 
identical 12-site clusters) via the Dyson equation and obtain 
the values of $\Omega[\Sigma(h')]$ for various values of $h'$ 
by using the Eq.~(\ref{eq:omega}).  
\section{\label{sec:Results}Results of calculations}

\subsection{Stability of the antiferromagnetic ordering}

\begin{figure}[htbp]
\begin{center}
\resizebox{8.5cm}{!}{\includegraphics{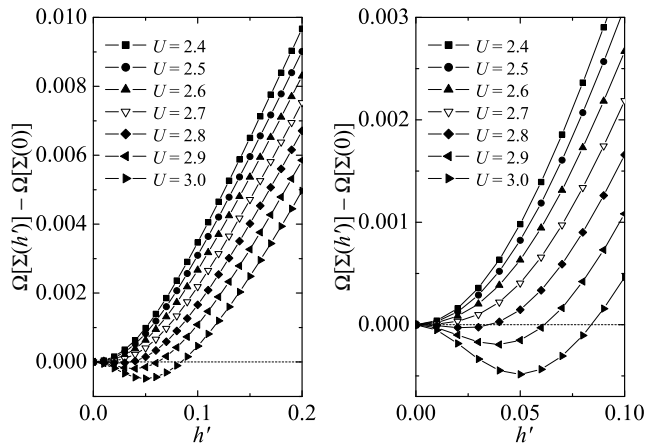}}\\
\caption{Calculated results for 
$\Omega[\Sigma(h')]-\Omega[\Sigma(0)]$ (per site).  
The right panel shows an enlargement of the small $h'$ 
region of the left panel.  We show the results for several 
values of $U$ near the phase transition.  
Dotted horizontal line is a guide for eyes.}	
\label{critical}
\end{center}		
\end{figure}

It is known that the AF ordering of the $f$ electrons is 
realized in the ground state of the strong coupling region 
of PAM (as well as Kondo-lattice model) in 2D.\cite{vekic,assaad,shi}  
We demonstrate this in Fig.~\ref{critical}, where the 
calculated values of $\Omega[\Sigma(h')]-\Omega[\Sigma(0)]$ 
per site for several values of $U$ near the critical point 
are shown.  
We find the following: 
(i)~The value has a minimum at a finite value of $h'$ for 
$U>U_{\rm cr}$, which indicates that the symmetry-broken AF 
ordering is stabilized for $U>U_{\rm cr}$.  
(ii)~The value of $h'$ at which $\Omega[\Sigma(h')]-\Omega[\Sigma(0)]$ 
has a minimum approaches 0 with decreasing $U$ to 
$U\rightarrow U_{\rm cr}$.  
(iii)~The critical value of $U$ is determined as $U_{\rm cr}=2.7$.  
This value is comparable to (but is slightly smaller than) the 
result of the quantum Monte Carlo calculation\cite{vekic} where 
the value $U_{\rm cr}\simeq 2.95$ is reported.  
The reason of the overestimation of the AF stability in VCA may be 
explained as follows: In the VCA calculation, we use the cluster 
representable self-energies, i.e., exact self-energies of small 
clusters, as the trial self-energies.  
Thus, the long wave-length spin fluctuations beyond the cluster 
size are not taken into account.\cite{dahnken} 
Also, we use the staggered magnetic field on the $f$ orbitals $h'$ 
as a single variational parameter, which suppresses the quantum 
spin fluctuations.  
This may also be responsible for the overestimation.  
(iv)~For $U<U_{\rm cr}$, we find the minimum at $h'=0$, which 
indicates that there is no long-range AF ordering in the system.  

\subsection{Staggered magnetic moment}

\begin{figure}[htbp]
\begin{center}
\resizebox{6.0cm}{!}{\includegraphics{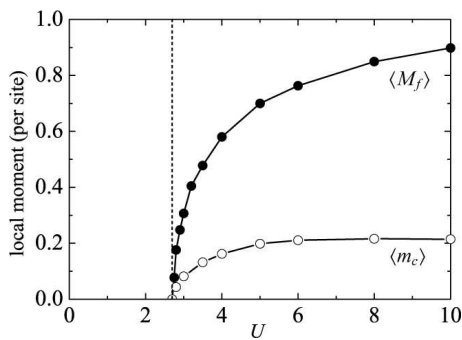}}\\
\caption{Calculated results for the staggered magnetic moment of 
the $f$ orbitals $\langle M_f\rangle$ 
and $c$ orbitals $\langle m_c\rangle$ as a function of $U$. 
Dotted line represents the critical value $U_{\rm cr}=2.7$. }	
\label{moment}
\end{center}
\end{figure}

In Fig.~\ref{moment}, we show calculated results for the 
staggered magnetic moment of the 
$f$ orbitals $\langle M_f\rangle$ and 
$c$ orbitals $\langle m_c\rangle$ per site 
at zero temperature, which are defined by 
\begin{subequations}
\begin{eqnarray}
\langle M_f\rangle &=&-\lim_{h_{\rm ext}^f\rightarrow 0}
\frac{\partial\Omega}{\partial h_{\rm ext}^f}, \\
\langle m_c\rangle &=&-\lim_{h_{\rm ext}^c\rightarrow 0}
\frac{\partial\Omega}{\partial h_{\rm ext}^c}
\end{eqnarray}
\end{subequations}
where $\Omega$ is the grand potential of the system per site 
and $h_{\rm ext}^f$ ($h_{\rm ext}^c$) is the 
external staggered magnetic field acting on the $f$ ($c$) orbitals.  
We find the following: 
(i)~The staggered magnetic moment of the $f$ electrons decreases 
continuously to 0 when we decrease the value of $U$ from the strong 
coupling region to $U_{\rm cr}$.  Thus, the phase transition 
is of the second order.   
(ii)~The staggered moment of the $c$ electrons also shows the 
similar behavior but the polarization is of the opposite sign.  
(iii)~The obtained staggered moments may be overestimated 
as in the case of the Hubbard model in 2D,\cite{dahnken} which is 
due again to the overestimation of the stability of the AF ordered 
state in VCA as discussed above.  

We also point out that the calculated staggered moments 
$\langle M_f\rangle$ and $\langle m_c\rangle$ are found to show 
a power-low behavior in the vicinity of the transition point as 
$\sim(U-U_{\rm cr})^\beta$ with the exponent $\beta\simeq 0.5$, 
which is consistent with the mean-field value $\beta=0.5$ within 
the numerical accuracy.  
Thus, the VCA calculation, which neglects the long-range spin 
fluctuations beyond the cluster size, gives the results equivalent 
to those of the simple mean-field approximation at least in 
the description of the critical behaviors such as the critical 
exponent.  

\subsection{Single-particle spectra and densities of states}

\begin{figure*}[htbp]
\begin{center}
\resizebox{15.5cm}{!}{\includegraphics{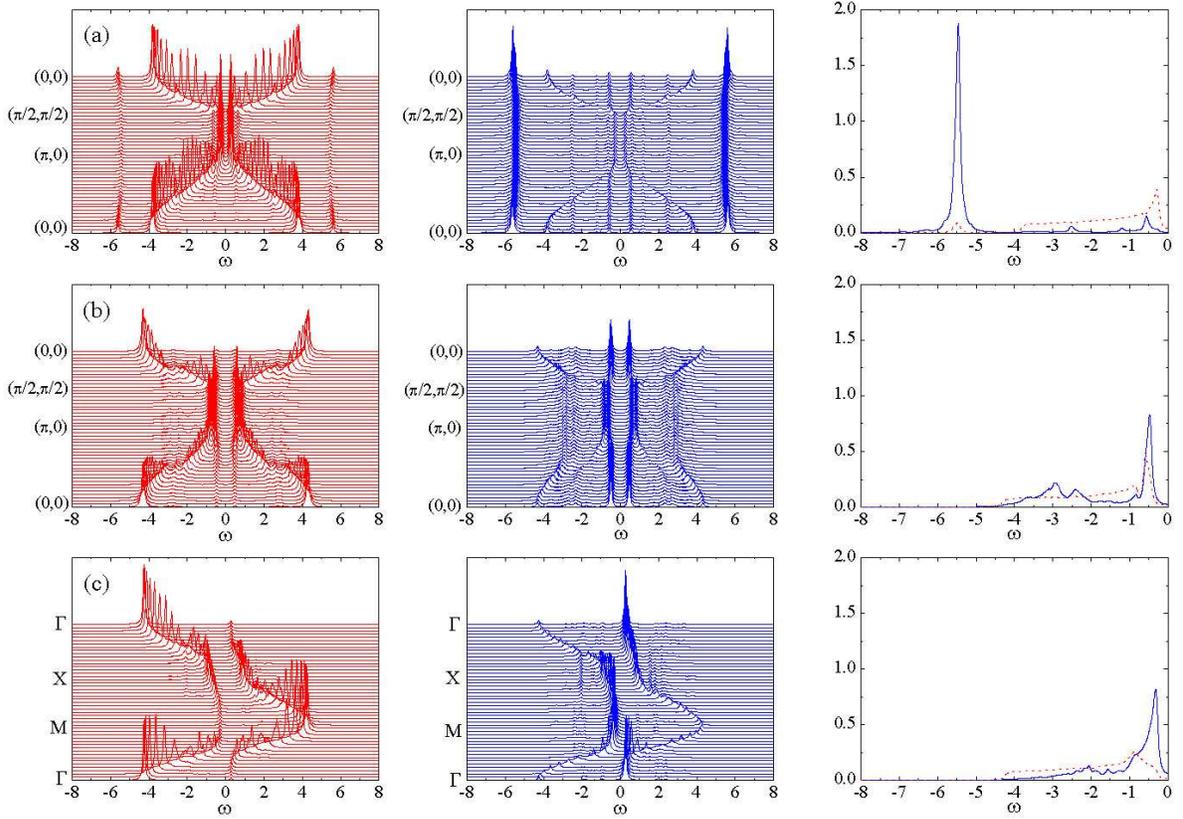}}\\
\caption{(Color online) Calculated results for the single-particle 
spectra (left and middle panels) and DOS (right panels) for 
(a)~$U=10$, (b)~$U=4$, and (c)~$U=2$, where $\omega=0$ corresponds 
to the Fermi energy.  The left and middle panels show the spectra 
of the $c$ and $f$ electrons, respectively.  In the right panel, 
the solid and dotted curves are the DOS for the $f$ and $c$ 
orbitals, respectively.  The artificial Lorentzian broadening 
of $\eta=0.05$ is included.}	
\label{spectra}
\end{center}		
\end{figure*}

We then calculate the single-particle spectra\cite{senechal} 
defined by the imaginary part of the ``Fourier transform'' 
of the optimized physical Green function.\cite{dahnken} 
We also calculate the DOS from the ${\bm k}$-space integration 
of the imaginary part of the optimized physical Green function.  
The results are shown in Fig.~\ref{spectra}, where 
we use the AF Brillouin zone when the system is in the 
symmetry-broken AF state (see Figs.~\ref{spectra}~(a) and (b)), 
but for nonmagnetic states, we use the standard first 
Brillouin zone (see Fig.~\ref{spectra}~(c)).  

In Fig.~\ref{spectra}~(a), i.e., for $U=10$, we can first 
identify the ``upper and lower Hubbard bands'' for the $f$ 
electrons, which are almost dispersionless and are separated 
by an energy $\sim$$U$.  
We can also identify the lower-energy dispersive bands in 
Fig.~\ref{spectra}~(a).  Here, we use the spin-density-wave 
(SDW) dispersion to fit the spectra.  The SDW dispersion can 
be obtained by diagonalizing the SDW Hamiltonian $H_{\rm SDW}$ 
defined by 
\begin{eqnarray}
&&H_{\rm SDW}=\sum_{{\bm k}\sigma}
\left(\begin{array}{cccc}
c_{{\rm A}{\bm k}\sigma}^{\dagger}
&c_{{\rm B}{\bm k}\sigma}^{\dagger}
&f_{{\rm A}{\bm k}\sigma}^{\dagger}
&f_{{\rm B}{\bm k}\sigma}^{\dagger}
\end{array}\right)\nonumber \\
&\times&\left(\begin{array}{cccc}
\sigma m_c&\varepsilon_{\bm k}& -\tilde{V} &0\\
\varepsilon_{\bm k}&-\sigma m_c&0&-\tilde{V}\\
-\tilde{V}&0&\tilde{E}_f-\sigma M_f&0\\
0&-\tilde{V}&0&\tilde{E}_f+\sigma M_f	
\end{array}\right)
\left(\begin{array}{c}
c_{{\rm A}{\bm k}\sigma}\\
c_{{\rm B}{\bm k}\sigma}\\
f_{{\rm A}{\bm k}\sigma}\\
f_{{\rm B}{\bm k}\sigma}
\end{array}\right),\nonumber \\
\end{eqnarray}
where A and B are the sublattice indices, 
$\tilde{V}$ and $\tilde{E}_f$ are the effective hybridization parameter 
and effective energy level of the $f$ orbital, respectively, 
$M_f$ ($m_c$) is the staggered magnetic moment of the $f$ ($c$) orbitals, 
and $\varepsilon_{\bm k}=-2t(\cos k_x+\cos k_y)$.  
We assume $M_f$ and $m_c$ to have the values obtained 
in Eqs.~(6a) and (6b) and we fix $\tilde{E}_f$ to be 0.  
We determine the value of $\tilde{V}$ so as to reproduce 
the size of the SDW gap.  We find that the fitting works well 
for the dispersions of the lower-energy bands but the spectral 
weight on the $f$ orbital differs very much from that of the VCA 
calculations since the upper and lower Hubbard bands for the $f$ 
electrons do not appear in the SDW spectral functions.  
We then find the value $\tilde{V}\simeq 0.35$ from the fitting, 
indicating that the quasiparticle is not quite heavy.  
In other words, with increasing $U$, the AF ordering 
occurs in 2D before the quasiparticle mass is strongly 
enhanced.  

In Fig.~\ref{spectra}~(b), i.e., for $U=4$, we find that the 
localized energy level is not well defined but there 
is a band repulsion in the spectra at $\varepsilon_f=\pm U/2$.  
The spectral weight of the $f$ electrons becomes large near 
the Fermi energy for all the momenta. 
Also, by comparing the results of the non-interacting case ($U=0$), 
the sharp peak just below the Fermi energy is observed 
in the partial DOS of the $f$ orbital (see the right panel of 
Fig.~\ref{spectra}~(b)).  
Thus, we conclude that this peak not only arises from the hybridization 
but is caused by the many-body resonance, which corresponds 
to the Kondo resonance peak in the metallic state.  

In Fig.~\ref{spectra}~(c), i.e., for $U=2$, where there is no 
AF ordering in the system, we find that the spectra look very 
similar to the spectra of non-interacting case.  
However, we again find that the localized energy level 
is not well defined but there is a weak band repulsion in the 
spectra at $\varepsilon_f=\pm U/2$.  

\subsection{Charge gap and Spin gap}

\begin{figure}[htbp]
\begin{center}
\resizebox{4.5cm}{!}{\includegraphics{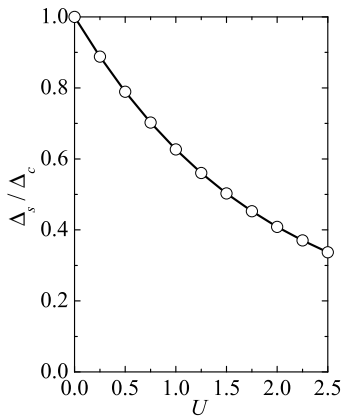}}\\
\caption{Calculated result for the ratio of the spin gap 
to charge gap $\Delta_s/\Delta_c$ as a function of $U$ $(U<U_{\rm cr})$.  
The 8-site cluster is used.}	
\label{gap}
\end{center}		
\end{figure}

To clarify the behavior in the weak-coupling region where there 
is no AF ordering, i.e., $U<U_{\rm cr}$, we calculate the spin 
and charge gaps defined as 
$\Delta_s=E_0(N_{\uparrow}+1,N_{\downarrow}-1)
-E_0(N_{\uparrow},N_{\downarrow})$ and 
$\Delta_c=[E_0(N_{\uparrow}+1,N_{\downarrow}+1)
+E_0(N_{\uparrow}-1,N_{\downarrow}-1)
-2E_0(N_{\uparrow},N_{\downarrow})]/2$, 
respectively, where $E_0(N_{\uparrow},N_{\downarrow})$ is 
the ground-state energy of a cluster with $N_{\uparrow}$ 
up-spin and $N_{\downarrow}$ down-spin electrons.  
Because the two-particle Green functions cannot be calculated 
directly from VCA, we here use an exact-diagonalization technique 
on small clusters.  We use the 8-site, 16-orbital cluster with 
periodic boundary condition to calculate the ground-state energies 
and estimate the spin and charge gaps.  
In Fig.~\ref{gap}, we show the ratio of the spin gap to 
the charge gap $\Delta_s/\Delta_c$ thus obtained as a function 
of $U$, where the result only at $U<U_{\rm cr}$ (with $U_{\rm cr}$ 
determined in Sec.~III A) is shown because no phase transition 
occurs in finite-size systems.  We find $\Delta_c>\Delta_s$ 
for all values of $U$ $(<U_{\rm cr})$, indicating the system 
to be in the regime of the Kondo insulator;\cite{vekic} i.e., 
there is no long-range AF ordering, where localized spins 
are screened by the formation of the Kondo singlet state.  
As $U$ decreases to 0, we find that the two gaps tend smoothly 
to the same value, i.e., $\Delta_s/\Delta_c\rightarrow 1$, 
indicating the system tends to the non-interacting band 
insulator.  

\section{\label{sec:Summary}Summary}

In summary, we have applied the VCA based on the SFT for 
the first time to consider the symmetric PAM at half-filling 
in 2D.  We have thus demonstrated the validity of the approach 
by discussing in particular the competition between 
antiferromagnetism and Kondo screening in the thermodynamic 
limit at zero temperature.  
We have shown that the symmetry-broken AF ordering of 
localized spins is realized in the strong coupling 
region $U>U_{\rm cr}$ and the Kondo insulating behavior 
is realized in the weak coupling region $U<U_{\rm cr}$.  
We have determined the critical interaction strength as 
$U_{\rm cr}=2.7$.  We have calculated the staggered 
magnetic moment as a function of the interaction strength 
and have shown that the phase transition is of the second order. 
We have also calculated the single-particle spectra and 
density of states. 
We have thereby discussed the effect of electron correlations 
on the quasiparticle band structure.  We have applied an 
exact-diagonalization technique on small clusters to 
calculate the ratio of the spin gap to charge gap 
in the weak coupling region and 
found that the Kondo insulating state continuously 
tends to the non-interacting band insulator with 
decreasing the value of $U$ to 0.  

We thus have shown that the present approach is very 
useful for considering the electronic states of PAM 
in 2D.  To improve the accuracy of our results, one 
may introduce additional variational parameters,  
such as the hopping terms, to take into account the 
quantum fluctuations more efficiently and suppress the 
overestimation of the stability of the AF ordering, 
which we want to leave for future studies.  

\begin{acknowledgments}
This work was supported in part by Grants-in-Aid for 
Scientific Research (Nos. 18028008, 18043006, 18540338, 
and 19014004) from the Ministry of Education, Culture, 
Sports, Science and Technology of Japan.  
A part of computations was carried out at the Research 
Center for Computational Science, Okazaki Research 
Facilities, and the Institute for Solid State Physics, 
University of Tokyo.  
\end{acknowledgments}

\end{document}